\documentclass{article}
\usepackage{amssymb}
\usepackage{bm}
\usepackage[dvips]{graphicx}
\begin{document}

\title{Spin-polarized current effect on antiferromagnet magnetization
in a ferromagnet--antiferromagnet nanojunction: Theory and simulation}

\author{E. M. Epshtein\thanks{E-mail: epshtein36@mail.ru}, Yu. V. Gulyaev, P. E. Zilberman
\\ \\
V. A. Kotelnikov Institute of Radio Engineering and Electronics\\
of the Russian Academy of Sciences, Fryazino, 141190, Russia}

\date{}

\maketitle

\abstract{Spin-polarized current effect is studied on the static and dynamic magnetization
of the antiferromagnet in a ferromagnet--antiferromagnet nanojunction. The
macrospin approximation is generalized to antiferromagnets. Canted
antiferromagnetic configuration and resulting magnetic moment are induced
by an external magnetic field. The resonance frequency and damping are
calculated, as well as the threshold current density corresponding to
instability appearance. A possibility is shown of generating low-damping
magnetization oscillations in terahertz range. The fluctuation effect is discussed on
the canted antiferromagnetic configuration. Numerical simulation is
carried out of the magnetization dynamics of the antiferromagnetic layer in the nanojunction
with spin-polarized current. Outside the instability range, the simulation
results coincide completely with analytical calculations using linear
approximation. In the instability range, undamped oscillations occur of the
longitudinal and transverse magnetization components.}\\
\\

\section{Introduction}\label{section1}
The discovery of the spin transfer torque effect in ferromagnetic
junctions under spin-polarized current~\cite{Slonczewski,Berger} has
stimulated a number of works in which such effects were observed as
switching the junction magnetic configuration~\cite{Katine}, spin wave
generation~\cite{Tsoi}, current-driven motion of magnetic domain
walls~\cite{Yamaguchi}, modification of ferromagnetic
resonance~\cite{Sankey}, etc. It is well known that the spin torque transfer
from spin-polarized electrons to lattice leads to appearance of a negative
damping. At some current density, this negative damping overcomes the
positive (Gilbert) damping with occurring instability of the original
magnetic configuration. The corresponding current density is high enough,
of the order of $10^7$ A/cm$^2$. This, naturally, stimulates attempts to
lower this threshold. Various ways were proposed, such as using magnetic
semiconductors~\cite{Watanabe}, in which the threshold current density can
be lower down to $10^5$--$10^6$ A/cm$^2$ because of their low saturation
magnetization. However, using of such materials requires, as a rule, low
temperatures because of low Curie temperature. Besides, the ferromagnetic
resonance frequency is rather low in this case.

In connection with these difficulties, the other approaches were proposed,
based on high spin injection~\cite{Gulyaev1} or joint action of external
magnetic field and spin-polarized current~\cite{Gulyaev2,Gulyaev3}. It
seems promising, also, using magnetic junction of
ferromagnet--antiferromagnet type, in which the ferromagnet (FM) acts as
an injector of spin-polarized electrons. The antiferromagnetic (AFM) layer, in
which the magnetic sublattices are canted by external magnetic field, may
have very low magnetization that promotes low threshold~\cite{Gulyaev4}.
The AFM resonance frequency may be both low and high reaching~$10^{12}$
s$^{-1}$, i.e. terahertz (THz) range. However, investigation and application of
THz resonances is prevented because of their large damping. Such a damping in
ferromagnetic junctions can be suppressed, as mentioned above, by means of
spin-polarized current. The question arises about possibility of such a
suppression in FM/AFM junctions. Note, that this problem has been paid
attention of a number of authors~\cite{Sankaranarayanan}--\cite{Swaving}.

Another interesting feature of the FM/AFM junctions with spin-polarized
current is the possibility of canting the AFM structure by spin-polarized current
without magnetic field.

\section{The equations of motion}\label{section2}
\begin{figure}
\includegraphics{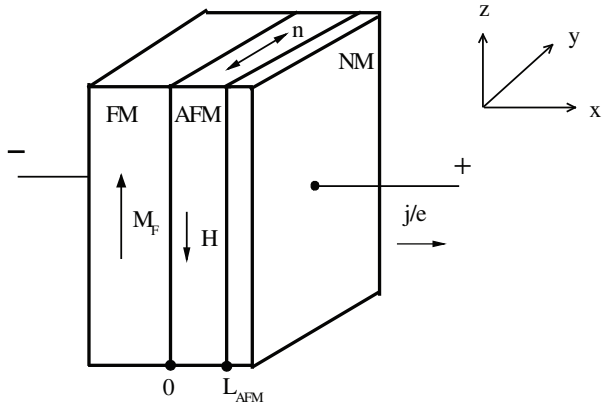}
  \caption{Scheme of the ferromagnet (FM)--antiferromagnet (AFM)
  junction; NM being a nonmagnetic layer. The main vector directions are
  shown.}\label{fig1}
\end{figure}

Let us consider a FM/AFM junction (Fig.~\ref{fig1}) with current flowing perpendicular to
layers, along $x$ axis. An ultrathin spacer layer is placed between the FM and AFM layers to
prevent direct exchange coupling between the layers.
An external magnetic field is parallel to the FM
magnetization and lies in the layer plane $yz$. The simplest AFM model is
used with two equivalent sublattices.

The AFM energy (per unit area), with uniform and nonuniform exchange,
anisotropy, external magnetic field, and the \emph{sd} exchange interaction
of the conduction electrons with the magnetic lattice taking into
account, takes the form~\cite{Akhiezer}
\begin{eqnarray}\label{1}
  &&W=\int_0^{L_{AFM}}\,dx\biggl\{\Lambda(\mathbf{M}_1\cdot\mathbf{M}_2)
  +\frac{1}{2}\alpha\left\{\left(\frac{\partial\mathbf{M}_1}{\partial
  x}\right)^2+\left(\frac{\partial\mathbf{M}_2}{\partial
  x}\right)^2\right\} \nonumber \\
  &&+\alpha'\left(\frac{\partial\mathbf{M}_1}{\partial
  x}\cdot\frac{\partial\mathbf{M}_2}{\partial
  x}\right)-\frac{1}{2}\beta\left\{(\mathbf{M}_1\cdot\mathbf
  n)^2+(\mathbf{M}_2\cdot\mathbf
  n)^2\right\}\nonumber \\
  &&-\beta'(\mathbf{M}_1\cdot\mathbf n) (\mathbf{M}_2\cdot\mathbf
  n)-((\mathbf{M}_1+\mathbf{M}_2)\cdot\mathbf H)\nonumber \\
  &&-\alpha_{sd}((\mathbf{M}_1
  +\mathbf{M}_2)\cdot\mathbf m)\biggr\},
\end{eqnarray}
where $\mathbf{M}_1,\,\mathbf{M}_2$ are the sublattice magnetization
vectors, $\Lambda$ is the uniform exchange constant, $\alpha,\,\alpha'$
are the intra- and inter-sublattice nonuniform exchange constants,
respectively, $\beta,\,\beta'$ are the corresponding anisotropy constants,
$\mathbf n$ is the unit vector along the anisotropy axis, $\mathbf H$ is
the external magnetic field, $\mathbf m$ is the conduction electron
magnetization, $\alpha_{sd}$ is the dimensionless \emph{sd} exchange
interaction constant. We do not include demagnetization term because its
contribution is small compared to the uniform exchange.
The integral is taken over the AFM layer thickness $L_{AFM}$. We are
interested in the spin-polarized current effect on the AFM layer, so we
consider a case of perfect FM injector with pinned lattice magnetization
and without disturbance of the electron spin equilibrium, that allows
to not include the FM layer energy in Eq.~(\ref{1}).

Two mechanisms are known of the spin-polarized current effect on the
magnetic lattice, namely, spin transfer torque (STT)~\cite{Slonczewski,Berger}
and an alternative mechanism~\cite{Heide,Gulyaev5} due to the spin
injection and appearance of nonequilibrium population of the spin subbands
in the collector layer (this is the AFM layer, in our case). In the case of
antiparallel relative orientation of the injector and collector
magnetization vectors, such a state becomes energetically
unfavorable at high enough current density, so
that the antiparallel configuration switches to parallel one (such a
process in FM junction is considered in detail in review~\cite{Gulyaev6}).
The latter mechanism is described with the \emph{sd} exchange term in
Eq.~(\ref{1}). As to the former mechanism, it is of dissipative character
(it leads to negative damping), so that it is taken into account by the
boundary conditions (see below), not the Hamiltonian.

The equations of the sublattice motion with damping taking into account
take the form
\begin{equation}\label{2}
    \frac{\partial\mathbf{M}_i}{\partial t}-\frac{\kappa}{M_0}
    \left[\mathbf{M}_i\times\frac{\partial\mathbf{M}_i}{\partial
    t}\right]+\gamma\left[\mathbf{M}_i\times\mathbf{H}_{eff}^{(i)}\right]
    =0\quad(i=1,\,2),
\end{equation}
where $M_0$ is the sublattice magnetization, $\kappa$ is the damping
constant,
\begin{equation}\label{3}
    \mathbf{H}_{eff}^{(i)}=-\frac{\delta W}{\delta\mathbf{M}_i}\quad(i=1,\,2)
\end{equation}
are the effective fields acting on the corresponding sublattices.

From Eqs.~(\ref{1})--(\ref{3}) the equations are obtained for the total
magnetization $\mathbf M=\mathbf{M}_1+\mathbf{M}_2$ and antiferromagnetism
vector $\mathbf L=\mathbf{M}_1-\mathbf{M}_2$:
\begin{eqnarray}\label{4}
  &&\frac{\partial\mathbf M}{\partial t}-\frac{1}{2}\frac{\kappa}{M_0}
    \left\{\left[\mathbf M\times\frac{\partial\mathbf M}{\partial
    t}\right]+\left[\mathbf L\times\frac{\partial\mathbf L}{\partial
    t}\right]\right\} \nonumber \\
    &&+\gamma\left[\mathbf M\times\mathbf H\right]
    +\gamma\left[\mathbf M\times\mathbf{H}_{sd}\right] \nonumber \\
    &&+\frac{1}{2}\gamma(\beta+\beta')(\mathbf M\cdot\mathbf n)[\mathbf M\times\mathbf
    n]+\frac{1}{2}\gamma(\beta-\beta')(\mathbf L\cdot\mathbf n)[\mathbf L\times\mathbf
    n] \nonumber \\
    &&+\frac{1}{2}\gamma(\alpha+\alpha')\left[\mathbf M\times
    \frac{\partial^2\mathbf M}{\partial
    x^2}\right]+\frac{1}{2}\gamma(\alpha-\alpha')
    \left[\mathbf L\times\frac{\partial^2\mathbf L}{\partial
    x^2}\right]=0,
\end{eqnarray}

\begin{eqnarray}\label{5}
  &&\frac{\partial\mathbf L}{\partial t}-\frac{1}{2}\frac{\kappa}{M_0}
    \left\{\left[\mathbf L\times\frac{\partial\mathbf M}{\partial
    t}\right]+\left[\mathbf M\times\frac{\partial\mathbf L}{\partial
    t}\right]\right\} \nonumber \\
    &&+\gamma\left[\mathbf L\times\mathbf H\right]
    +\gamma\left[\mathbf L\times\mathbf{H}_{sd}\right]
    -\gamma\Lambda\left[\mathbf L\times\mathbf M\right] \nonumber \\
    &&+\frac{1}{2}\gamma(\beta+\beta')(\mathbf M\cdot\mathbf n)[\mathbf L\times\mathbf
    n]+\frac{1}{2}\gamma(\beta-\beta')(\mathbf L\cdot\mathbf n)[\mathbf M\times\mathbf
    n] \nonumber \\
    &&+\frac{1}{2}\gamma(\alpha+\alpha')\left[\mathbf L\times
    \frac{\partial^2\mathbf M}{\partial
    x^2}\right]+\frac{1}{2}\gamma(\alpha-\alpha')
    \left[\mathbf M\times\frac{\partial^2\mathbf L}{\partial
    x^2}\right]=0,
\end{eqnarray}
where
\begin{equation}\label{6}
    \mathbf{H}_{sd}(x)=\frac{\delta}{\delta\mathbf
    M(x)}\int_0^{L_{AFM}}\left(\mathbf M(x')\cdot\mathbf m(x')\right)\,dx'
\end{equation}
is the effective field due to \emph{sd} exchange interaction. This field
determines the spin injection contribution to the interaction of the
conduction electrons with the antiferromagnet lattice.

To find $\mathbf{H}_{sd}(x)$ field, the conduction electron
magnetization $\mathbf m(x)$ is to be calculated. The details of such
calculations are presented in our preceding
papers~\cite{Gulyaev7,Gulyaev2}. Here we adduce the result for the case,
where the antiferromagnet layer thickness $L_{AFM}$ is small compared to
the spin diffusion length $l$ with the current flow direction
corresponding to the electron flux from FM to AFM:
\begin{equation}\label{7}
  \mathbf m=(\overline m+\Delta m)\hat{\mathbf M},
  \quad\Delta m=\frac{\mu_B\tau Qj}{eL_{AFM}}\left(\hat{\mathbf M}(0)\cdot\hat{\mathbf
  M}_F\right),
\end{equation}
where $\overline m$ is the equilibrium (in absence of current) electron
magnetization, $\Delta m$ is the nonequilibrium increment due to current,
$\hat{\mathbf M}=\mathbf M/|\mathbf M|$ is the unit vector along the AFM
magnetization, $\hat{\mathbf M}_F$ is the similar vector for
FM, $\mu_B$ is the Bohr magneton, $e$ is the electron charge, $\tau$ is
the electron spin relaxation time, $j$ is the current density.

It should have in mind in varying the integral~(\ref{6}), that the
electron magnetization $\mathbf m$ depends on the vector $\mathbf M$
orientation relative to the FM magnetization vector $\mathbf M_F$. From
Eqs.~(\ref{6}) and~(\ref{7}) we have~\cite{Gulyaev2}
\begin{eqnarray}\label{8}
  &\mathbf{H}_{sd}=\alpha_{sd}\overline m\hat{\mathbf M}+
    \alpha_{sd}\displaystyle\frac{\mu_B\tau Qj}{eL_{AFM}}\left(\hat{\mathbf M}(0)\cdot\hat{\mathbf
  M}_F\right)\hat{\mathbf M}\nonumber\\
    &+\alpha_{sd}\displaystyle\frac{\mu_B\tau Qj}{e}\hat{\mathbf M}_F\delta(x-0).
\end{eqnarray}

By substitution~(\ref{8}) into~(\ref{4}) and~(\ref{5}), we obtain
\begin{eqnarray}\label{9}
  &&\frac{\partial\mathbf M}{\partial t}-\frac{1}{2}\frac{\kappa}{M_0}
    \left\{\left[\mathbf M\times\frac{\partial\mathbf M}{\partial
    t}\right]+\left[\mathbf L\times\frac{\partial\mathbf L}{\partial
    t}\right]\right\} \nonumber \\
    &&+\gamma\left[\mathbf M\times\mathbf H\right]
    +\gamma\alpha_{sd}\frac{\mu_B\tau Qj}{e}\left[\mathbf M\times
    \hat{\mathbf M}_F\right]\delta(x-0) \nonumber \\
    &&+\frac{1}{2}\gamma(\beta+\beta')(\mathbf M\cdot\mathbf n)[\mathbf M\times\mathbf
    n]+\frac{1}{2}\gamma(\beta-\beta')(\mathbf L\cdot\mathbf n)[\mathbf L\times\mathbf
    n] \nonumber \\
    &&+\frac{1}{2}\gamma(\alpha+\alpha')\left[\mathbf M\times
    \frac{\partial^2\mathbf M}{\partial
    x^2}\right]+\frac{1}{2}\gamma(\alpha-\alpha')
    \left[\mathbf L\times\frac{\partial^2\mathbf L}{\partial
    x^2}\right]=0,
\end{eqnarray}

\begin{eqnarray}\label{10}
  &&\frac{\partial\mathbf L}{\partial t}-\frac{1}{2}\frac{\kappa}{M_0}
    \left\{\left[\mathbf L\times\frac{\partial\mathbf M}{\partial
    t}\right]+\left[\mathbf M\times\frac{\partial\mathbf L}{\partial
    t}\right]\right\} \nonumber \\
    &&+\gamma\left[\mathbf L\times\mathbf H\right]
    +\gamma\alpha_{sd}\frac{\mu_B\tau Qj}{e}\left[\mathbf L\times
    \hat{\mathbf M}_F\right]\delta(x-0) \nonumber \\
    &&-\gamma\left(\Lambda-\frac{\alpha_{sd}\overline m}{M}
    -\frac{\alpha_{sd}\mu_B\tau Qj}{eL_{AFM}M}\left(\hat{\mathbf M}(0)\cdot\hat{\mathbf
  M}_F\right)\right)\left[\mathbf L\times\mathbf M\right] \nonumber \\
    &&+\frac{1}{2}\gamma(\beta+\beta')(\mathbf M\cdot\mathbf n)[\mathbf L\times\mathbf
    n]+\frac{1}{2}\gamma(\beta-\beta')(\mathbf L\cdot\mathbf n)[\mathbf M\times\mathbf
    n] \nonumber \\
    &&+\frac{1}{2}\gamma(\alpha+\alpha')\left[\mathbf L\times
    \frac{\partial^2\mathbf M}{\partial
    x^2}\right]+\frac{1}{2}\gamma(\alpha-\alpha')
    \left[\mathbf M\times\frac{\partial^2\mathbf L}{\partial
    x^2}\right]=0.
\end{eqnarray}

\section{The boundary conditions}\label{section3}
The equations of motion~(\ref{9}) and~(\ref{10}) contain derivative over
the space coordinate $x$. Therefore, boundary conditions at the
AFM layer surfaces $x=0$ and $x=L_{AFM}$ are need to find solutions. The
way of derivation was described in Ref.~\cite{Gulyaev2} in detail. The
conditions depend on the electron spin polarization and are determined
by the continuity requirement of the spin currents at the interfaces.

The terms with the space derivative in Eq.~(\ref{9}) may be written in the
form of a divergency:
\begin{eqnarray}\label{11}
  &&\frac{1}{2}\gamma(\alpha+\alpha')\left[\mathbf M\times
    \frac{\partial^2\mathbf M}{\partial
    x^2}\right]+\frac{1}{2}\gamma(\alpha-\alpha')
    \left[\mathbf L\times\frac{\partial^2\mathbf L}{\partial
    x^2}\right] \nonumber \\
  &&=\frac{\partial}{\partial x}\left\{\frac{1}{2}\gamma(\alpha+\alpha')\left[\mathbf M\times
    \frac{\partial\mathbf M}{\partial
    x}\right]+\frac{1}{2}\gamma(\alpha-\alpha')
    \left[\mathbf L\times\frac{\partial\mathbf L}{\partial
    x}\right]\right\} \nonumber \\
    &&\equiv\frac{\partial\mathbf{J}_M}{\partial x}.
\end{eqnarray}

The $\mathbf{J}_M$ vector is the lattice magnetization flux density.

Let us integrate Eq.~(\ref{9}) over $x$ within narrow interval
$0<x<\varepsilon$ with subsequent passing to $\varepsilon\to+0$ limit.
Then only the mentioned terms with the space derivative and the singular
term with delta function will contribute to the integral. As a result, we
obtain an effective magnetization flux density with \emph{sd} exchange contribution at
the AFM boundary $x=+0$ taking into account:
\begin{equation}\label{12}
  \mathbf{J}_{eff}(+0)=\mathbf{J}_M(+0)
  +\gamma\alpha_{sd}\frac{\mu_B\tau
  Qj}{e}\left[\mathbf{M}(+0)\times\hat{\mathbf M}_F\right].
\end{equation}

The magnetization flux density coming from the FM injector is
\begin{equation}\label{13}
    \mathbf J(-0)=\frac{\mu_BQ}{e}j\hat{\mathbf M}_F.
\end{equation}
The component $\mathbf J_{\|}=\left(\mathbf J(-0)\cdot\hat{\mathbf M}(+0)\right)\hat{\mathbf
M}(+0)$ remains with the electrons, while the rest,
\begin{eqnarray}\label{14}
    &&\mathbf J_\bot=\mathbf J(-0)-\mathbf J_{\|}=\frac{\mu_B  Q}{e}j
    \left\{\hat{\mathbf M}_F-\hat{\mathbf M}(+0)\left(\hat{\mathbf
  M}_F\cdot\hat{\mathbf M}(+0)\right)\right\} \nonumber \\
  &&=-\frac{\mu_BQ}{eM^2}j\left[\mathbf{M}(+0)\times\left[\mathbf{M}(+0)
  \times\hat{\mathbf M}_F\right]\right],
\end{eqnarray}
is transferred to the AFM lattice owing to conservation of the
magnetization fluxes~\cite{Slonczewski,Berger}.

By equating the magnetization fluxes~(\ref{12}) and~(\ref{14}), we obtain
\begin{equation}\label{15}
    \mathbf{J}_M=-\frac{\mu_BQ}{eM^2}j\left[\mathbf M\times\left[\mathbf M
  \times\hat{\mathbf M}_F\right]\right]-\gamma\alpha_{sd}\frac{\mu_B\tau
  Q}{e}j\left[\mathbf M\times\hat{\mathbf M}_F\right],
\end{equation}
all the $\mathbf M$ vectors being taken at $x=+0$.

Since the AFM layer thickness is small compared to the spin diffusion
length and the exchange length, we may use the macrospin approximation
which was described in detail in Ref.~\cite{Gulyaev2}. In this
approximation, the magnetization changes slowly within the layer
thickness. This allows to write
\begin{equation}\label{16}
  \frac{\partial\mathbf J_M}{\partial x}\approx
  \frac{\mathbf J_M(L_{AFM})-\mathbf J_M(+0)}{L_{AFM}}=-\frac{\mathbf
  J_M(+0)}{L_{AFM}},
\end{equation}
because the magnetization flux is equal to zero at the interface between
AFM and the nonmagnetic layer closing the electric circuit, $\mathbf
J_M(L_{AFM})=0$. This allows to exclude the terms with space derivative
from Eq.~(\ref{9}). In the rest terms, $\mathbf M(x,\,t)$ and $\mathbf
L(x,\,t)$ quantities are replaced with their values at $x=0$. Then
Eq.~(\ref{9}) takes a more simple form:
\begin{eqnarray}\label{17}
  &&\frac{d\mathbf M}{dt}-\frac{1}{2}\frac{\kappa}{M_0}
    \left\{\left[\mathbf M\times\frac{d\mathbf M}{dt}\right]
    +\left[\mathbf L\times\frac{d\mathbf L}{dt}\right]\right\}
    +\gamma\left[\mathbf M\times\mathbf H\right]\nonumber \\
    &&+\frac{1}{2}\gamma(\beta+\beta')(\mathbf M\cdot\mathbf n)[\mathbf M\times\mathbf
    n]+\frac{1}{2}\gamma(\beta-\beta')(\mathbf L\cdot\mathbf n)[\mathbf L\times\mathbf
    n]\nonumber \\
    &&+K\left[\mathbf M\times\left[\mathbf M\times\hat{\mathbf
    M}_F\right]\right]+P\left[\mathbf M\times\hat{\mathbf
    M}_F\right]=0,
\end{eqnarray}
where
\begin{equation}\label{18}
  K=\frac{\mu_BQ}{eL_{AFM}M^2}j,\qquad
  P=\frac{\gamma\alpha_{sd}\mu_B\tau Q}{eL_{AFM}}j.
\end{equation}
The term with delta function does not present here, since it is taken into
account in the boundary conditions.

Now we are to use again the macrospin approximation to exclude the space
derivatives from Eq.~(\ref{10}), too.

Owing to known relationships~\cite{Akhiezer} between $\mathbf M$ and $\mathbf
L$ vectors, namely, $M^2+L^2=4M_0^2$ and $(\mathbf M\cdot\mathbf L)=0$,
we have the following conditions:
\begin{equation}\label{19}
  \left(\mathbf M\cdot\frac{\partial\mathbf M}{\partial t}\right)+
  \left(\mathbf L\cdot\frac{\partial\mathbf L}{\partial t}\right)=0,\quad
  \left(\mathbf L\cdot\frac{\partial\mathbf M}{\partial t}\right)+
  \left(\mathbf M\cdot\frac{\partial\mathbf L}{\partial t}\right)=0.
\end{equation}

By substituting Eqs.~(\ref{10}) and~(\ref{17}) in~(\ref{19}) we find that
conditions~(\ref{19}) are fulfilled if the terms in~(\ref{10})
\begin{equation}\label{20}
    \frac{1}{2}\gamma(\alpha+\alpha')\left[\mathbf L\times
    \frac{\partial^2\mathbf M}{\partial
    x^2}\right]+\frac{1}{2}\gamma(\alpha-\alpha')
    \left[\mathbf M\times\frac{\partial^2\mathbf L}{\partial
    x^2}\right]\equiv\mathbf X
\end{equation}
satisfy the following equations:
\begin{eqnarray}\label{21}
  &&(\mathbf X\cdot\mathbf M)+K\left(\mathbf L\cdot\left[\mathbf M
  \times\left[\mathbf M\times\hat{\mathbf M}_F\right]\right]\right)
  +P\left(\mathbf L\cdot\left[\mathbf M
  \times\hat{\mathbf M}_F\right]\right)=0, \nonumber \\
  &&(\mathbf X\cdot\mathbf L)=0.
\end{eqnarray}

Let us decompose the considered $\mathbf X$ vector on three mutually
orthogonal vectors:
\begin{equation}\label{22}
    \mathbf X=a\mathbf M+b\mathbf L+c\gamma\left[\mathbf L\times\mathbf
    M\right].
\end{equation}
The substitution~(\ref{22}) in~(\ref{21}) gives $a=K\left(\mathbf
L\cdot\hat{\mathbf M}_F\right)-P\left(\left[\mathbf
L\times\mathbf M\right]\cdot\hat{\mathbf M}_F\right)$, $b=0$. As to $c$
coefficient, it is a current-induced correction to the coefficient of
$\gamma[\mathbf L\times\mathbf M]$ term in Eq.~(\ref{10}), i.\,e., a
correction to the uniform exchange constant $\Lambda$. Let us estimate the
correction. Multiplying~(\ref{22}) scalarly by $[\mathbf L\times\mathbf
M]$ with~(\ref{20}) taking into account gives
\begin{eqnarray}\label{23}
    &&c=\frac{1}{M^2L^2}\left([\mathbf L\times\mathbf M]\cdot
    \left\{\frac{1}{2}\gamma(\alpha+\alpha')\left[\mathbf L\times
    \frac{\partial^2\mathbf M}{\partial
    x^2}\right]+\frac{1}{2}\gamma(\alpha-\alpha')
    \left[\mathbf M\times\frac{\partial^2\mathbf L}{\partial
    x^2}\right]\right\}\right) \nonumber \\
    &&=\frac{1}{2}\left\{(\alpha+\alpha')\frac{1}{M^2}
    \left(\mathbf M\cdot\frac{\partial^2\mathbf M}{\partial
    x^2}\right)-(\alpha-\alpha')\frac{1}{L^2}
    \left(\mathbf L\cdot\frac{\partial^2\mathbf L}{\partial
    x^2}\right)\right\}.
\end{eqnarray}
It is seen that $c\sim\alpha/L_{AFM}^2$, while $\Lambda\sim\alpha/a^2$,
where $a$ is the lattice constant~\cite{Akhiezer}. Since $L_{AFM}\gg a$,
the mentioned correction to $\Lambda$ may be neglected.

As a result, Eq.~(\ref{10}) takes the form
\begin{eqnarray}\label{24}
  &&\frac{d\mathbf L}{dt}-\frac{1}{2}\frac{\kappa}{M_0}
    \left\{\left[\mathbf L\times\frac{d\mathbf M}{dt}\right]
    +\left[\mathbf M\times\frac{d\mathbf L}{dt}\right]\right\} \nonumber \\
    &&+\gamma\left[\mathbf L\times\mathbf H\right]
    -\gamma\Lambda\left[\mathbf L\times\mathbf M\right] \nonumber \\
    &&+\frac{1}{2}\gamma(\beta+\beta')(\mathbf M\cdot\mathbf n)[\mathbf L\times\mathbf
    n]+\frac{1}{2}\gamma(\beta-\beta')(\mathbf L\cdot\mathbf n)[\mathbf M\times\mathbf
    n] \nonumber \\
    &&+K\left[\mathbf L\times\left[\mathbf M\times\hat{\mathbf
    M}_F\right]\right]+P\left[\mathbf L\times\hat{\mathbf M}_F\right]=0.
\end{eqnarray}
Here, $\Lambda$ constant contains also the equilibrium contribution of the
conduction electrons $-\alpha_{sd}\overline m/M$.

Equations~(\ref{17}) and~(\ref{24}) are the result of applying the macrospin
concept to AFM. It is shown that such an approximation
may be justified formally for AFM layer. Earlier, it was justified for FM
layers~\cite{Slonczewski,Berger} and generalized~\cite{Gulyaev2} with spin
injection taking into account. The macrospin approach corresponds well to
experimental conditions and simplifies calculations substantially. The
terms with $K$ coefficient in Eqs.~(\ref{17}), (\ref{24}) describe effect
of STT mechanism, while the terms with $P$ coefficient take the spin
injection effect into account.

\section{The magnetization wave spectrum and damping}\label{section4}
We assume that the easy anisotropy axis lies in the plane of AFM layer and
is directed along $y$ axis, the FM magnetization vector is parallel to the
positive direction of $z$ axis, the external magnetic field is parallel to
$z$ axis too (see Fig.~\ref{fig1}).

We are interesting in behavior of small fluctuations around the steady
state $\mathbf M=\{0,\,0,\,\overline M_z\}$, $\mathbf L=\{0,\,\overline
L_y,\,0\}$, i.\,e. the small quantities $M_x,\,M_y,\,\widetilde
M_z=M_z-\overline M_z,\,L_x,\,\widetilde L_y=L_y-\overline L_y,\,L_z$.

Let us project Eqs.~(\ref{17}), (\ref{24}) to the coordinate axes
and take the terms up to the first order. The zero order terms are present
only in the projection of Eq.~(\ref{24}) to $x$ axis. They give
\begin{eqnarray}\label{25}
  &&\overline M_z=\frac{H_z+\displaystyle\frac{P}{\gamma}}{\Lambda+\displaystyle
  \frac{1}{2}(\beta-\beta')}
  \approx\frac{H_z+\displaystyle\frac{P}{\gamma}}{\Lambda}, \nonumber \\
  &&\overline L_y=\pm\sqrt{4M_0^2-\overline M_z^2}\approx\pm2M_0.
\end{eqnarray}

Note that the spin-polarized current takes part in creating magnetic
moment together with the external magnetic field due to the spin injection
induced interaction of the electron spins with the
lattice~\cite{Heide,Gulyaev5}, which $P$ parameter in Eq.~(\ref{25})
corresponds to. Such an interaction leads to appearance of an effective
magnetic field parallel to the injector magnetization. As a result, a
canted antiferromagnet configuration may be created without magnetic field.
However, such a configuration corresponds to parallel orientation of FM
and AFM layers, $\mathbf M\|\mathbf M_F$. As is shown below, the
instability does not occur with this orientation, so that an external
magnetic field is to be applied to reach instability.

With Eq.~(\ref{25}) taking into account, the equations for the first order
quantities take the form
\begin{eqnarray}\label{26}
  &&\frac{dM_x}{dt}-\frac{1}{2}\frac{\kappa}{M_0}
  \left\{-\overline M_z\frac{dM_y}{dt}+
  \overline L_y\frac{dL_z}{dt}\right\}+(\gamma H_z+P)M_y
  \nonumber \\
  &&-\frac{1}{2}\gamma(\beta+\beta')\overline M_zM_y
  -\frac{1}{2}\gamma(\beta-\beta')\overline L_yL_z+K\overline M_zM_x=0,
\end{eqnarray}

\begin{equation}\label{27}
  \frac{dM_y}{dt}-\frac{1}{2}\frac{\kappa}{M_0}
  \overline M_z\frac{dM_x}{dt}-
  (\gamma H_z+P)M_x+K\overline M_zM_y=0,
\end{equation}

\begin{equation}\label{28}
  \frac{d\widetilde M_z}{dt}+\frac{1}{2}\frac{\kappa}{M_0}
  \overline L_y\frac{dL_x}{dt}+\frac{1}{2}\gamma(\beta-\beta')\overline L_yL_x=0,
\end{equation}

\begin{equation}\label{29}
  \frac{dL_x}{dt}-\frac{1}{2}\frac{\kappa}{M_0}
  \left\{\overline L_y\frac{d\widetilde M_z}{dt}-
  \overline M_z\frac{d\widetilde L_y}{dt}\right\}-
  \gamma H_z\frac{\overline L_y}{\overline M_z}\widetilde M_z=0,
\end{equation}

\begin{equation}\label{30}
  \frac{d\widetilde L_y}{dt}-\frac{1}{2}\frac{\kappa}{M_0}
  \overline M_z\frac{dL_x}{dt}-\frac{1}{2}\gamma(\beta-\beta')\overline M_zL_x=0,
\end{equation}

\begin{equation}\label{31}
  \frac{dL_z}{dt}+\frac{1}{2}\frac{\kappa}{M_0}
  \overline L_y\frac{dM_x}{dt}+
  (\gamma H_z+P)\frac{\overline L_y}{\overline M_z}M_x
  +K\overline M_zL_z=0.
\end{equation}

The set of equations~(\ref{26})--(\ref{31}) splits up to two mutually
independent sets with respect to $(M_x,\,M_y,\,L_z)$ and
$(L_x,\,\widetilde L_y,\,\widetilde M_z)$. They describe two independent
spectral modes, one of them corresponds to precession of the AFM
magnetization vector around the magnetic field, while another to periodic
changes of the vector length along the magnetic field. We begin with the
spectrum and damping of the first mode. We consider monochromatic
oscillation with $\omega$ angular frequency and put
$M_x,\,M_y,\,L_z\sim\exp(-i\omega t)$. Then we obtain from
Eqs.~(\ref{26}),~(\ref{27}),~(\ref{31})
\begin{eqnarray}\label{32}
  &&\left(-i\omega+K\overline M_z\right)M_x+\left\{\gamma
  H_z+P-\frac{1}{2}\gamma(\beta+\beta')\overline M_z-
  \frac{1}{2}\frac{i\kappa\omega}{M_0}\overline M_z\right\}M_y \nonumber
  \\ &&-\left\{\frac{1}{2}\gamma(\beta-\beta')-
  \frac{1}{2}\frac{i\kappa\omega}{M_0}\right\}\overline L_yL_z=0,
\end{eqnarray}

\begin{equation}\label{33}
  \left(-i\omega+K\overline M_z\right)M_y-\left\{\gamma
  H_z+P-\frac{1}{2}\frac{i\kappa\omega}{M_0}\overline M_z\right\}M_x=0,
\end{equation}

\begin{equation}\label{34}
  \left(-i\omega+K\overline M_z\right)L_z+\left\{\gamma\Lambda+
  \frac{1}{2}\gamma(\beta-\beta')-
  \frac{1}{2}\frac{i\kappa\omega}{M_0}\right\}\overline L_y
M_x=0.
\end{equation}

Note that aforementioned additivity (in the algebraic sense, the sign
taking into account) of the external magnetic field and the
injection-driven effective field takes place not only in the steady
magnetization~(\ref{25}), but also in the oscillations of the
magnetization and antiferromagnetism vectors, so that both fields appear
in Eqs.~(\ref{32}),~(\ref{33}) ``on an equal footing''.

Usually, $\Lambda\gg\beta,\,\beta'$. With these inequalities and
stationary solution~(\ref{25}) taking into account we find the dispersion
relation for the magnetization oscillation
\begin{equation}\label{35}
  (1+\kappa^2)\omega^2+2i\nu\omega-\omega_0^2=0,
\end{equation}
where
\begin{equation}\label{36}
  \omega_0=\sqrt{2\gamma^2H_AH_E+(K\overline M_z)^2+(\gamma H_z+P)^2},
\end{equation}

\begin{equation}\label{37}
  \nu=\kappa\gamma H_E+K\overline M_z,
\end{equation}
$H_E=\Lambda M_0$ is the exchange field, $H_A=(\beta-\beta')M_0$ is the
anisotropy field. Formulae~(\ref{36}) and~(\ref{37}) (without current
terms $K\overline M_z$ and $P$) coincide with known
ones~\cite{Akhiezer,Gurevich}. At $H_E\sim10^6$--$10^7$ Oe, $H_A\sim10^3$
Oe we have oscillations in THz range, $\omega_0\sim10^{12}$ s$^{-1}$. In
absence of current the damping is rather high: at $\kappa\sim10^{-2}$
\begin{equation}\label{38}
  \frac{\nu}{\omega_0}=\kappa\sqrt{\frac{H_E}{2H_A}}\sim1.
\end{equation}

Let us consider the contribution of spin-polarized current to the
frequency and damping of AFM resonance. At first we consider STT mechanism
effect~\cite{Slonczewski,Berger}. According to~(\ref{18}) and~(\ref{25}),
\begin{equation}\label{39}
  K\overline M_z=\frac{\mu_BQ\Lambda}{eL_{AFM}H_z}j.
\end{equation}

At $H_z<0$, that corresponds to direction of the magnetic field (and,
therefore, the AFM magnetization) opposite to the FM magnetization, this
quantity is negative. The total attenuation becomes negative also (an
instability occurs), if
\begin{equation}\label{40}
  j>\frac{e\kappa\gamma M_0|H_z|L_{AFM}}{\mu_BQ}\equiv j_0.
\end{equation}
At $\kappa\sim10^{-2}$, $\gamma M_0\sim10^{10}$ s$^{-1}$, $|H_z|\sim10^2$
Oe, $L_{AFM}\sim10^{-6}$ cm, $Q\sim1$ we have $j_0\sim10^5$ A/cm$^2$. At
$j$ near to $j_0$ weakly damping THz oscillation can be obtained. At
$j>j_0$, instability occurs which may lead to either self-sustained
oscillations, or a dynamic stationary state. The latter
disappears with the current turning off. To answer the question about
future of the instability it is necessary to go out the scope of the
linear approximation. We have simulated numerically the behavior of the
AFM magnetization behind the linear approximation (see section~\ref{section8} below).

The spin-polarized current contributes also to the oscillation frequency.
At the mentioned parameter values, we have $|K\overline M_z|\sim10^{12}$
s$^{-1}$ that is comparable with the frequency in absence of the current.
This allows tuning the frequency by the current or excite parametric
resonance by means of the current modulation.

\section{Current-induced spin injection effect}\label{section5}
Now let us discuss the injection mechanism effect~\cite{Heide,Gulyaev5}.
As mentioned before, the role of the mechanism is reduced to addition of an
effective field $P/\gamma$ to the external magnetic field. At reasonable
parameter values, that field is much less than the exchange field $H_E$, so
that it does not influence directly the eigenfrequency~(\ref{36}).
Nevertheless, that field can modify substantially the contribution of the
STT mechanism, because Eq.~(\ref{39}) with~(\ref{25}) taking into account
now takes the form
\begin{equation}\label{41}
  K\overline M_z=\frac{\mu_BQ\Lambda}{eL_{AFM}(H_z+P/\gamma)}j.
\end{equation}
Such a modification leads to substantial consequences. At $H_z<0$,
$P<\gamma|H_z|$ the instability threshold~(\ref{40}) is lowered, since
$|H_z|-P/\gamma$ difference appears now instead of $|H_z|$. If, however,
$P>\gamma|H_z|$ then the AFM magnetization steady state
\begin{equation}\label{42}
    \overline M_z=\frac{H_z+P/\gamma}{\Lambda}
\end{equation}
becomes positive that corresponds to the parallel (stable) relative
orientation of the FM and AFM layers. In this case, the turning on current
leads to switching the antiparallel configuration (stated beforehand by
means of an external magnetic field) to parallel one. With turning off
current, the antiparallel configuration restores.

Since the mentioned injection-driven field depends on the current
(see~(\ref{18})), the instability condition~(\ref{40}) is modified and takes
the form
\begin{equation}\label{43}
  \frac{j_0}{1+\eta}<j<\frac{j_0}{\eta},
\end{equation}
where $\eta=\alpha_{sd}\kappa\gamma M_0\tau$, $j_0$ being defined with
Eq.~(\ref{40}). In absence of the injection mechanism, this condition
reduces to~(\ref{40}). Under rising role of this mechanism we have lowering the instability
threshold, on the one hand, and the instability range narrowing, on the
other hand. At $j>j_0/\eta$ the antiparallel configuration switches to
parallel one. The relative contribution of the injection mechanism is
determined with $\eta$ parameter. At typical values,
$\alpha_{sd}\sim10^4$, $\kappa\sim10^{-2}$, $\gamma M_0\sim10^{10}$
s$^{-1}$, $\tau\sim10^{-12}$ s, this parameter is of the order of unity,
so that the injection effect may lower noticeably the instability
threshold.

Now let us return to the set of equations~(\ref{26})--(\ref{31}) and
consider the second mode describing with Eqs.~(\ref{28})--(\ref{30}). The
current influences this mode by changing steady magnetization $\overline
M_z$ due to the injection effective field effect (see~(\ref{26})), while
the STT mechanism does not influence this mode.
    A calculation similar to previous one gives the former dispersion
relation~(\ref{35}), but now
\begin{equation}\label{44}
  \omega_0^2=2\gamma^2H_EH_A\frac{\gamma H_z}{\gamma H_z+P},
\end{equation}

\begin{equation}\label{45}
  \nu=\kappa\gamma H_E\frac{\gamma H_z}{\gamma H_z+P}.
\end{equation}

At $H_z<0$, $P>|H_z|$, that corresponds to current density $j>j_0/\eta$,
the total attenuation becomes negative, while the frequency becomes
imaginary, that means switching the antiparallel configuration to parallel
one.

\section{Easy plane type antiferromagnet}\label{section6}
Let us consider briefly the situation where AFM has easy-plane anisotropy.
We take the AFM layer $yz$ plane as the easy plane and $x$ axis as the
(hard) anisotropy axis. The magnetic field, as before, is directed along
$z$ axis.

Without repeating calculations, similar to previous ones, we present the results. A
formal difference appears only in Eq.~(\ref{36}) for the eigenfrequency
$\omega_0$ of the first of the modes considered above. We have for that
frequency
\begin{equation}\label{46}
  \omega_0=\sqrt{(\gamma H_z+P)^2+(K\overline M_z)^2}.
\end{equation}
The damping has the former form~(\ref{37}), so that the instability
threshold is determined with former formula~(\ref{43}).

In absence of the current ($K=0,\,P=0$) with not too small damping
coefficient $\kappa$, the frequency appears to be much less than damping,
so that the corresponding oscillations are not observed. The current
effect increases the frequency, on the one hand, and decreases the
damping (at $H_z<0$), on the other hand, that allows to observe
oscillation regime.

\section{Fluctuation effect}\label{section7}
It follows from Eq.~(\ref{43}) that the threshold current density is
proportional to the external magnetic field strength $|H_z|$ and decreases
with the field lowering. A question arises about permissible lowest limit of the
total field $|H_z|+P/\gamma$. In accordance with Eq.~(\ref{25}), such a
limit may be the field which create magnetization $|\overline M_z|$
comparable with its equilibrium value due to thermal fluctuations. Let us
estimate this magnetization and the corresponding field.

The AFM energy change in $V$ volume under canting the sublattice
magnetization vectors with $\theta<180^\circ$ angle between them is
\begin{equation}\label{47}
  \Delta E=\Lambda M_0^2(1-\cos\theta)V=\frac{1}{2}\Lambda VM_z^2,
\end{equation}
the anisotropy energy being neglected compared to the exchange energy.

The equilibrium value of the squared magnetization is calculated using the
Gibbs distribution:
\begin{equation}\label{48}
  \langle M_z^2\rangle=\displaystyle\frac{\int\limits_{-\infty}^\infty
  M_z^2\exp\left(-\displaystyle\frac{\Lambda VM_z^2}{2kT}\right)\,dM_z}
  {\int\limits_{-\infty}^\infty
  \exp\left(-\displaystyle\frac{\Lambda VM_z^2}{2kT}\right)\,dM_z}
  =\frac{kT}{\Lambda V}
\end{equation}
(strictly speaking, the magnetization may be changed within
$(-2M_0,\,2M_0)$ interval, however, $\Lambda VM_0^2\gg kT$, so that the
integration limits may be taken infinity).

To observe the effects described above, the magnetization $\overline M_z$
which appears under joint action of the external field and the current
(see~(\ref{25})) should exceed in magnitude the equilibrium magnetization
$\langle M_z^2\rangle^{1/2}$. At the current density $j=j_0/(1+\eta)$
corresponding to the instability threshold, this condition is fulfilled at
magnetic field
\begin{equation}\label{49}
  |H_z|>\sqrt{\frac{\Lambda kT}{V}}(1+\eta)\equiv H_{min}.
\end{equation}
At $\Lambda\sim10^4$, $\eta\sim1$, $L_{AFM}\sim10^{-6}$ cm and lateral
sizes of the switched element $10\times10\,\mu$m$^2$ we have
$V\sim10^{-12}$ cm$^3$ and $H_{min}\approx30$ Oe at room temperature. This
limit can be decreased under larger element size.

It should be mentioned also about other mechanisms of AFM canting. The
most known and studied one is the relativistic Dzyaloshinskii--Moria effect (see,
e.g.~\cite{Akhiezer,Dmitrienko}). Besides, possible mechanisms have been
discussed due to competition between \emph{sd} exchange and direct
exchange interaction of the magnetic ions in the lattice~\cite{Robinson}.
At the same time, there are no indications, to our knowledge, about
measurements of canting in conductive AFM. So, present theory is related
to conductive AFM, in which the lattice canting is determined with
external magnetic field.

\section{Simulation}\label{section8}
With the purpose of simulating, it is convenient to modify slightly
Eqs.~(\ref{17}) and~(\ref{24}), namely: i) to use sublattice
magnetizations $\mathbf M_1,\,\mathbf M_2$ instead of $\mathbf M,\,\mathbf
L$, ii) to describe damping by the Landau--Lifshitz representation with
double vector product, and iii) to introduce the following dimensionless
variables:
\begin{eqnarray*}
  &\hat{\mathbf M}_i=\mathbf M_i/M_0\;(i=1,\,2),\quad\mathbf h=\mathbf
  H/M_0,\quad T=\displaystyle\frac{\gamma M_0t}{1+\kappa^2},\\
  &K_0=\displaystyle\frac{\mu_BQj}{eL_{AFM}\gamma M_0^2},\quad
  P_0=\eta_0K_0,\quad\eta_0=\alpha_{sd}\gamma M_0\tau.
\end{eqnarray*}
With these variables, the set of equations~(\ref{17}), (\ref{24}) take
the form
\begin{equation}\label{50}
  \frac{d\hat{\mathbf M}_i}{dT}=-\left[\hat{\mathbf M}_i\times\mathbf
  h_{eff}^{(i)}\right]-\kappa\left[\hat{\mathbf M}_i\times\left[\hat{\mathbf M}_i
  \times\mathbf h_{eff}^{(i)}\right]\right],\quad i=1,\,2,
\end{equation}
\begin{eqnarray}\label{51}
    &&\mathbf h_{eff}^{(1)}=\mathbf h-\Lambda\hat{\mathbf M}_2+\beta\left(\hat{\mathbf
    M}_1\cdot\mathbf n\right)\mathbf n+\beta'\left(\hat{\mathbf
    M}_2\cdot\mathbf n\right)\mathbf n+P_0\hat{\mathbf M}_F\nonumber\\
    &&+K_0\displaystyle\frac{\left[\left(\hat{\mathbf M}_1+\hat{\mathbf M}_2\right)\times\hat{\mathbf
    M}_F\right]}{\left(\hat{\mathbf M}_1+\hat{\mathbf
    M}_2\right)^2}\equiv\mathbf F\left(\hat{\mathbf M}_1,\,\hat{\mathbf
    M}_2\right),
\end{eqnarray}

\begin{equation}\label{52}
  \mathbf{h}_{eff}^{(2)}=\mathbf{F}\left(\hat{\mathbf M}_2,\,\hat{\mathbf
    M}_1\right).
\end{equation}
It is seen from Eqs.~(\ref{50})--~(\ref{52}) that the motions of two
sublattices are coupled each other. The sources of such a coupling are the
uniform exchange, intersublattice anisotropy (coefficient~$\beta'$), as well
the effective field due to the spin torque (the last term in
Eq.~(\ref{51})).

We assume the following orders of magnitude of the used
parameters:~$\kappa\sim10^{-3}$--$10^{-2}$, $\Lambda\sim10^3$--$10^4$,
$\beta,\,\beta'\sim10^{-1}\;(\beta\ne\beta')$, $M_0\sim10^3$ G, $\gamma
M_0\sim10^{10} s^{-1}$. Under such conditions,~$h=1$ corresponds to
magnetic field $H\sim10^3$ Oe, and~$T=1$ corresponds to
time~$t\sim10^{-10}$ s. With~$\alpha_{sd}\sim10^4$, $\tau\sim10^{-12}$ s,
$Q\sim1$, $L_{AFM}\sim10^{-7}$ cm we have~$\eta_0\sim10^2$, and~$K_0=1$
value corresponds to current density~$j\sim10^7$ A/cm$^2$.

Equations~(\ref{50})--~(\ref{52}) written in coordinates represent six
ordinary differential equations of the first order in the Cauchy form
for~$\hat M_{1x}$, $\hat M_{1y}$, $\hat M_{1z}$, $\hat M_{2x}$, $\hat M_{2y}$, $\hat
M_{2z}$. The equations are not mutually independent because of the
normalization conditions
\begin{equation}\label{53}
  \left|\hat{\mathbf M}_1\right|=\left|\hat{\mathbf M}_2\right|=1.
\end{equation}
Nevertheless, all the six equations are used in our simulation, while the
mentioned normalization conditions serve for checking correctness of the
calculations.

The simulation was carried out by means of Simulink program in MATLAB
system with using Differential Equation Editor (DEE). Right-hand sides of
the equations resolved with respect to derivatives were entered into the DEE
block. The parameters~$\Lambda,\,\beta,\,\beta',\,\kappa,\,K_0,\,P_0$ were
given as input signals, while three projections of the magnetization
vector~$\hat M_x=\hat M_{1x}+\hat M_{2x},\;\hat M_y=\hat M_{1y}+\hat M_{2y},\;\hat M_z=\hat M_{1z}+\hat
M_{2z}$ were output to oscilloscope blocks. Besides,~$\left|\hat M_1\right|=\left(\hat
M_{1x}^2+\hat M_{1y}^2+\hat M_{1z}^2\right)^{1/2}$ and~$\left|\hat M_2\right|=\left(\hat
M_{2x}^2+\hat M_{2y}^2+\hat M_{2z}^2\right)^{1/2}$ values were output to
digital displays; these values must be equal to 1 (or, at least, be close
to 1) under correct calculation.

In the present work, we assume that the magnetic field and current are
turned on at the initial time instant~$T=0$ and hold constant. However,
the procedure used allows to consider arbitrary time dependence of these
quantities, specifically, to vary turning on and turning off instants.

We began simulation with ``verifying'' results of the linear theory
(actually, this was a test for the model adequacy). As above, we assume
that the AFM layer lies in~$yz$ plane, the current flows along~$x$ axis,
the easy anisotropy axis coincides with~$y$ axis ($\mathbf
n=\{0,\,1,\,0\}$), the current is polarized along the positive direction
of the~$z$ axis ($\hat\mathbf M_F=\{0,\,0,\,1\}$), the magnetic field is
collinear with~$z$ axis ($\mathbf h=\{0,\,0,\,h_z\}$). In such a
configuration, the collinear relative orientation of the FM and AFM layer
magnetizations is stationary (although, possibly, unstable), so that a
small initial deviation from such orientation was given to imitate thermal
fluctuations. The initial value of the~$\hat M_z$ component was chosen to
be equal to the equilibrium value~$\hat M_z=h_z/\Lambda$ in the given
magnetic field without current. Thus, at~$\Lambda=10^3,\,h_z=-1$ initial
conditions~$\hat M_{1z}=\hat M_{2z}=-5\times10^{-4}$ (so that~$\hat M_z=-1\times10^{-3}$),
$\hat M_{1x}=\hat M_{2x}=-5\times10^{-6}$, $\hat M_{1y}=-\hat M_{2y}=(1-\hat M_{1x}^2-\hat
M_{1z}^2)^{1/2}$ are used taking the normalization conditions~(\ref{53}) into
account.

With the dimensionless variables, the main results of the linear theory
take the following form.

The stationary magnetization in~$z$ direction is
\begin{equation}\label{54}
  \hat M_z\equiv\hat M_{1z}+\hat
  M_{2z}=\frac{h_z+\eta_0K_0}{\Lambda+\frac{1}{2}(\beta-\beta')}\approx\frac{h_z+\eta_0K_0}{\Lambda}.
\end{equation}
Under deviation from the stationary value, two modes appear with
dimensionless (referred to~$\gamma M_0$) frequencies~$\omega_{1,\,2}$ and
damping~$\nu_{1,\,2}$ defined with the following formulae:
\begin{equation}\label{55}
  \omega_1^2=2(\beta-\beta')\Lambda+(h_z+\eta_0K_0)^2+\left(\frac{\Lambda
  K_0}{h_z+\eta_0K_0}\right)^2,
\end{equation}
\begin{equation}\label{56}
  \nu_1=\kappa\Lambda+\frac{\Lambda K_0}{h_z+\eta_0K_0},
\end{equation}
\begin{equation}\label{57}
    \omega_2^2=\frac{2(\beta-\beta')\Lambda h_z}{h_z+\eta_0K_0},
\end{equation}
\begin{equation}\label{58}
    \nu_2=\frac{\kappa\Lambda h_z}{h_z+\eta_0K_0}.
\end{equation}

The instability of the antiparallel relative orientation of the FM and AFM
layers at~$h_z<0$ occurs in the current density value range defined with
inequalities
\begin{equation}\label{59}
  \frac{\kappa|h_z|}{1+\kappa\eta_0}<K_0<\frac{|h_z|}{\eta_0}.
\end{equation}

To start, the case of absence of the current ($K_0=0$) has been
considered. Perfect agreement has been observed with
Eqs.~(\ref{55})--(\ref{58}). In particular, the magnetization oscillation
frequency drops abruptly when we put~$\beta=\beta'$, and the oscillations
disappear completely, if we take, moreover,~$h=0$. At~$\beta\ne\beta'$ the
observed frequency consists with Eq.~(\ref{55}). In absence of the
magnetic field, the frequencies of the two modes coincide, so that the
oscillation takes the form of a simple sinusoid. Under rising
magnetic field, the frequencies become different, and beats appear
because of interaction between the modes. The simulation results consist,
also, with Eq.~(\ref{54}) for the stationary magnetization in presence of
the magnetic field. Turning on the magnetic field at~$T=0$ instant leads to
an aperiodic transient process which decays completely by~$T=0.4$,
following which the magnetization remains constant value determined by
Eq.~(\ref{54}).

\begin{figure}
\includegraphics[width=120mm]{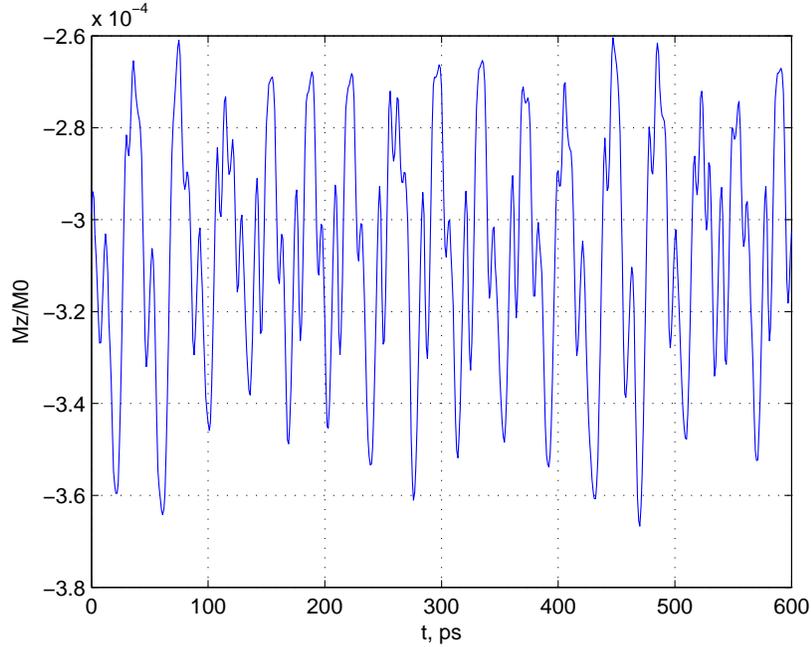}
  \caption{Oscillations of the longitudinal magnetization in the
  instability range at~$K_0=7\times10^{-3}$.}\label{fig2}
\end{figure}

At~$\Lambda=10^3,\,\kappa=10^{-2},\,\eta_0=10^2,\,h_z=-1$ the instability
predicted by the linear theory must occur at~$K_0=5\times10^{-3}$ and
disappear at~$K_0=1\times10^{-2}$. In our numerical experiments,
increasing~$K_0$ parameter from zero to the indicated threshold leads, in
accordance with Eq.~(\ref{56}), to decrease of the damping of the
magnetization vector precession about~$z$ axis because of the negative
damping caused by the STT mechanism. Incidentally, the absolute value of
the (negative)~$\hat M_z$ component decreases from~$10^{-3}$ to~$10^{-4}$
because of influence of the injection mechanism, that consists, also, with
Eq.~(\ref{54}). Thus, the simulation results agree completely with the
theory in the range below the instability threshold.

\begin{figure}
\includegraphics[width=110mm]{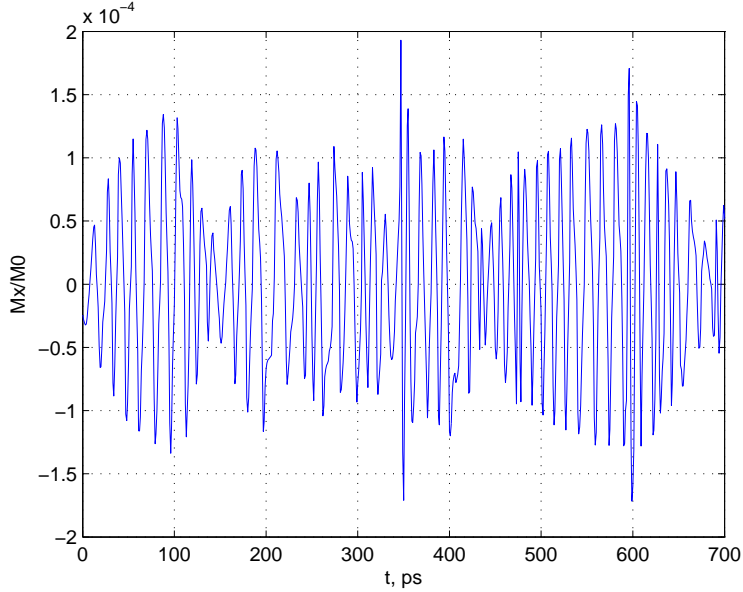}
  \caption{Beats of the transverse magnetization in the
  instability range at~$K_0=8\times10^{-3}$.}\label{fig3}
\end{figure}

Of course, the instability range ($K_0>5\times10^{-3}$), which is not
described by the linear theory, is of more interest.
At~$K_0=5.1\times10^{-3}$ undamped oscillations are observed.
At~$K_0=7\times10^{-3}$ the precession oscillations of the transverse
components~$\hat M_x,\,\hat M_y$ become almost sinusoidal with~$\Delta
T\approx0.2$ period (this corresponds to angular
frequency~$\sim3\times10^{11}$ s$^{-1}$ at the chosen parameter values).
The longitudinal component~$\hat M_z$ oscillates periodically with a
negative stationary background, however, the oscillation form is far from
sinusoidal one (Fig.~\ref{fig2}). At~$K_0=8\times10^{-3}$ the oscillations
of~$\hat M_x,\,\hat M_y$ components take on a form of beats
(Fig.~\ref{fig3}), while they again become sinusoidal
at~$K_0=9\times10^{-3}$. At~$K_0=1\times10^{-2}$ (this is the right
boundary of the instability range in the linear theory) all the three
components oscillate around zero values. Further,
at~$K_0=1.1\times10^{-2}$ the oscillations hold yet, but
at~$K_0=1.2\times10^{-2}$ they disappear almost completely, and they are
absent at~$K_0=1.5\times10^{-2}$.

The subsequent increasing of the current leads only to rising magnetization
in the positive direction due to the injection mechanism.
At~$K_0=2\Lambda/\eta_0$, the longitudinal component~$\hat M_z$ reaches
the maximal possible value~$\hat M_z=2$ (the sublattices are flipped to a parallel
position), then the (dimensionless) angular frequency is equal
to~$2\Lambda$ (this corresponds to~$\sim10^{13}$ s$^{-1}$). Such a
situation corresponds to rather high current density, it is estimated
as~$2\times10^8$ A/cm$^2$ at the above-mentioned parameter values.

\section{Conclusions}\label{section9}
The obtained results show a principal possibility of controlling frequency
and damping of AFM resonance in FM/AFM junctions by means of
spin-polarized current. Under low AFM magnetization induced by an external
magnetic field perpendicular to the antiferromagnetism vector, the
threshold current density corresponding to occurring instability is less
substantially than in the FM--FM case. Near the threshold, the AFM
resonance frequency increases, while damping decreases, that opens a
possibility of generating oscillations in THz range.

Numerical simulation allows to trace behavior of the
FM/AFM junction in the whole current density range.
The instability range predicted by the linearized theory is broadened only
slightly because of nonlinear effects. In the instability range undamped
oscillation of sinusoidal or more complicated form including beats.

Under magnetic fields low compared to the exchange field, the induced
magnetization is small in comparison with the sublattice magnetization, so
that the stationary oscillation amplitude beyond the instability threshold
is low, too.

Thus, the following features may be expected in comparison with the
similar effects in FM/FM junctions. First, the instability threshold is to
be lower because of the lower magnetization. This is a favorable fact
facilitating observations. Second, the oscillation intensity beyond the
threshold also lowers as a square of the magnetization. This may make
the effect difficult to observe. Nevertheless, studying the current-driven nonlinear
oscillations in FM/AFM structure is of principal interest, because the
current induced instability can occur at relatively low current
density,~$\sim10^5$ A/cm$^2$.

The simulation results reveal an interesting possibility of a spin-flip
transition without magnetic field under the action of a high-density
spin-polarized current only. Such a current overcomes the exchange forces
and aligns the sublattice moments in parallel. Under such conditions,
applying a low alternating magnetic field can excite precession of the
magnetization vector at the AFM resonance frequencies, which may be as
high as~$3\times10^{13}$ s$^{-1}$ or more. Such a THz resonator might be
useful to detect and measure signals in THz range.

\section*{Acknowledgments}
The authors are grateful to Prof. G.~M.~Mikhailov for useful discussions.

The work was supported by the Russian Foundation for Basic Research,
Grant No.~10-02-00030-a.

\end{document}